\documentclass[amsmath,amssymb,aps,twocolumn,prb]{revtex4}

\usepackage{graphicx}
\usepackage{dcolumn}
\usepackage{bm}
\usepackage{color}

\begin{document}

\widetext

\noindent \textbf{\Large 
Coexistence and efficiency of normal and anomalous
transport by molecular motors in living cells
}\\

\noindent \textbf{\large Igor Goychuk$^{1,*}$, Vasyl O. Kharchenko$^{1,2}$ \& 
Ralf Metzler$^{1,3}$} \\
 
\noindent \textbf{$^{1}$Institute for Physics and Astronomy, University of
Potsdam, Karl-Liebknecht-Str. 24/25, 14476 Potsdam-Golm, Germany,
$^{2}$Institute of Applied Physics, 58 Petropavlovskaya str., 40030 Sumy,
Ukraine,
$^3$Department of Physics, Tampere University of Technology, Korkeakoulunkatu 3,
33101 Tampere, Finland;
$^{*}$email: igoychuk@uni-potsdam.de, 
corresponding author}\\

\textbf{
Recent experiments reveal both passive subdiffusion of various nanoparticles 
and anomalous active
transport of such particles by molecular motors in the molecularly
crowded environment of living biological cells. Passive and active
microrheology reveals that the origin of this anomalous dynamics is due to
the viscoelasticity of the intracellular fluid. How do molecular motors
perform in such a highly viscous, dissipative environment? Can we explain the
observed co-existence of the anomalous transport of relatively large particles
of 100 to 500 nm in size by kinesin motors with the normal transport of
smaller particles by the same molecular motors? What is the efficiency
of molecular motors in the anomalous transport regime? Here we answer these
seemingly conflicting questions and consistently explain experimental findings
in a generalization of the well-known continuous diffusion model for molecular
motors with two conformational states in which viscoelastic effects are included.
}

\maketitle

After the publication of Albert Einstein's theory of Brownian motion in
1905\cite{einstein}, Jean Perrin reported the first systematic microscopic
studies of individual diffusing particles in 1908\cite{perrin}. Today, modern
single particle tracking techniques routinely
reveal insight into the stochastic motion
of submicron tracers in aqueous solutions at unprecedented resolution, thus
allowing one to directly observe the transition from initial ballistic to
diffusive Brownian motion\cite{huang} and to resolve the effects of hydrodynamic
backflow\cite{franosch}. Measuring the passive and driven motion of microprobes
has become a standard means to characterize soft matter\cite{Mason}. Particular
attention is currently paid to the relaxation and diffusion dynamics in dense
colloidal systems \cite{mattsson,weeks} and inside living biological cells
\cite{pt,Yamada}.

The intracellular fluid (cytosol) of biological cells is a
superdense\cite{golding} aqueous solution containing biomacromolecules such as
proteins and RNA at volume fractions of up to 40\%, a state often referred to
as macromolecular crowding\cite{ellis,mcguffee}. Indeed, the state of crowding
in the cytosol effects severe changes of the diffusion behavior of submicron
particles\cite{pt,Saxton,Hofling}. Thus, anomalous diffusion of the form $\langle[
\delta\mathbf{r}(t)]^2\rangle=\langle[\mathbf{r}(t+t')-\mathbf{r}(t')]^2\rangle
\simeq t^\beta$ with $0<\beta<1$ is observed in living cells
for the passive motion of single
biopolymers, endogenous granules, viruses, and artificial tracer particles
\cite{Tolic,Guigas,Taylor,Tabei,Caspi,Seisenberg,Wachsmuth}. Compared to
normal Brownian motion with $\beta=1$, these particles therefore
subdiffuse\cite{Metzler2000}. Such anomalous dynamics presents
a challenge to the development of controlled uptake of drugs and nanoparticles
and their intracellular delivery by molecular motors for therapeutic processes
\cite{drugs}.

\begin{figure}
\resizebox{0.8\columnwidth}{!}{\includegraphics{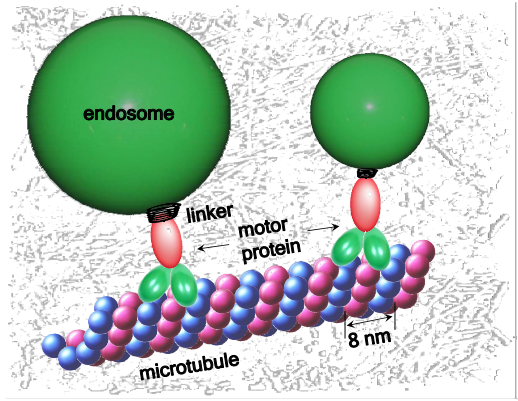}}
\caption{Molecular motors moving along a microtubule in the crowded cytoplasm.
A large cargo is subject to viscoelastic drag, effecting dramatic changes in the
transport dynamics.} 
\label{Fig1}      
\end{figure}

What happens to the active motion of particles in the cytosol which are driven by
molecular motors\cite{hirokawa,Julicher1,Julicher2,Astumian,Reimann,Fisher}, 
see Fig.~\ref{Fig1}? Will the overall dynamics of the coupled motor-cargo system
be affected by the superdense state of the cytosol, and how? In fact, anomalously
fast diffusion was observed in
living cells for the active motion of viruses, microbeads, and endosomes
\cite{Seisenberg,Caspi,Robert}. When the motors normally transport their
cargo with some mean velocity in a given direction, ballistic superdiffusion
with $\beta=2$ is measured\cite{Seisenberg}. However, various subballistic
power exponents $1<\beta<2$ are also found\cite{Caspi,Robert}.
Here we come up with
a novel theory for molecular motors in the cytosol of living cells and show
that our predictions are in agreement with the experimentally observed
behavior. The simplest explanation for the experimental fact
that
transport by motors along intracellular microtubuli yields subballistic superdiffusion 
($1<\beta<2$), is provided by the sublinear scaling $\langle\delta x(t)\rangle=
\langle x(t+t')-x(t')\rangle\propto t^{\alpha_{\rm eff}}$ of the mean position $x$
along a single microtubule, with an effective scaling exponent $0.5<\alpha_{\rm
eff}<1$. Then, assuming quenched disorder in microtubuli orientations
without a preferred direction yields
$\beta=2\alpha_{\rm eff}$ which will range in the subballistic
superdiffusive regime. 
Can normal and anomalous transport with various $\beta$
co-exist and be mediated by the \emph{same\/} motors in the \emph{same\/} cell,
as found in experiments\cite{Seisenberg,Caspi,Robert}?
Which role is played by the size of the cargo, and what determines the precise
behavior of such active transport and its efficiency? These are the questions we
answer in this article.

A well-established physical approach to anomalous transport phenomena is
based on the intrinsic viscoelasticity\cite{Mason,Amblard,Qian,Weigh,Goychuk12}
of complex fluids such as the cytosol. Depending on its size and speed, a
nanoparticle may feel different effective viscosities in the macroscopic limit
of normal diffusion\cite{Luby} which may be enhanced by a factor of hundred to
several thousand with respect to the viscosity of pure water. Transient regimes
of anomalous diffusion become biologically important when the corresponding
spatial length scale is
comparable to the linear size of cells, typically several
micrometers. The viscoelastic nature of the cytosol and other dense solutions
has been verified in several experiments\cite{Tolic,Guigas,Jeon,Szymanski}.

\begin{figure}
\includegraphics[width=6.8cm]{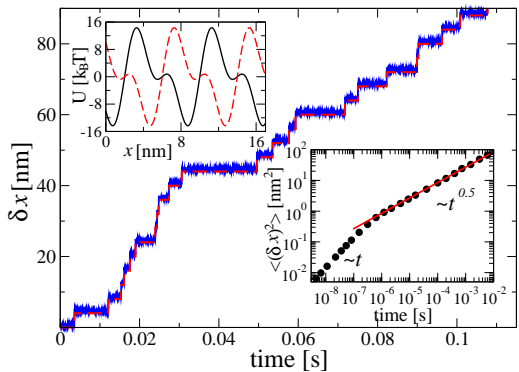}
\caption{Normal transport for large cargo particles,
large potential amplitude and small turnover rate, in the absence
a of constant loading force, $f_0=0$.
Single motor transport (full line) is almost perfectly locked to the
potential fluctuations (broken red line depicting a renewal process counting
the number of potential fluctuations in units of $L/2$) occurring with
mean turnover frequency $\nu_{\rm turn}=112$ Hz, in a potential (top inset) with
amplitudes $U_0=0.25$ eV ($U_0=1$ in dimensionless units) and $U_1=0.162$ eV 
($\Delta V=0.7$ eV), for $L=8$ nm. A particle with an effective radius $300$ nm
(like a magnetic endosome\cite{Robert}) experiences asymptotically for $t\gg
\tau_{\rm max}\approx 22.4$ sec an effective viscous friction enhanced by
a factor of $\tilde\eta_{\rm eff}=10^4$ with respect to water. The bottom
inset shows that on the relevant transient time scale the free particle
subdiffuses with anomalous diffusion coefficient $D_\alpha \approx 368\;
{\rm nm^2/s^{0.5}}$. Initially, the diffusion is normal. The time-average over
a single trajectory, $\langle\delta x^2(t)\rangle_{\cal T}=\frac{1}{{\cal
T}-t}\int_{0}^{{\cal T}-t}[x(t+t')-x(t')]^2 dt'$, is shown for
${\cal T}=0.1$ sec and compared with the theoretical subdiffusive
ensemble-averaged result (red line). See \textbf{Methods}.}
\label{Fig2}
\end{figure}

We study the interplay of the viscoelastic environment of the cytosol with the
action of a molecular motor and its cargo. A well-established model of Brownian
motors of the kinesin family is based on the
continuous diffusion of a Brownian
particle in a potential landscape, which randomly fluctuates in time between
two realizations, $V_1(x)\rightleftharpoons V_2(x)$, depending on the internal
state of the motor, that is undergoing active conformational
fluctuations\cite{Julicher1,Julicher2,Astumian,Reimann}. These conformational
fluctuations are caused by the binding of negatively charged ATP molecules to
the motor
(state 1) and the reactions of ATP hydrolysis and dissociation of products
(ADP and phosphate group) making up state 2 within a minimalist modeling
framework. The potentials $V_{1,2}(x)$ describe the free energy profiles leading
the motor molecule in the corresponding conformational states along
a microtubule. Since microtubules are periodic dipolar structures with period
$L\approx 8$~nm, the potentials reflect this periodicity, $V_{1,2}(x+L)=V_{1,2}
(x)$. Moreover, $V_{1}(x+L/2)=V_{2}(x)$ within the two-state motor model,
such that two potential switches occur during one cyclic turnover of the motor
enzyme (power-stroke or ``hand-over-hand'' mechanism) and advance the motor by
one period $L$. The motor direction is determined by the polarity of the
microtubules, reflected in the space inversion asymmetry of the potentials.
We use the harmonic mixing model\cite{Makhno} with $V_1(x)=-U_0\cos(2\pi x/L)-
U_1\sin(4\pi x/L)$, see the upper inset in Fig.~\ref{Fig2}.
This potential has a metastable state
within each period. Thermal fluctuations play a positive role, allowing the
motor to avoid getting trapped in such metastable states on its search for
the potential minimum after each conformational change due to ATP binding and
ATP hydrolysis. 

Within the power-stroke idealization, the maximal mean velocity
of the motor due to the fluctuations between the potentials $V_{1,2}(x)$ becomes
$v=L\nu_{\rm turn}$, where $\nu_{\rm turn}$ is the typical motor turnover rate.
The latter is composed of the conformational transition rates $\nu_{1,2}$
according to $\nu_{\rm turn}^{-1}=\nu_1^{-1}+\nu_2^{-1}$. We assume that these
transition rates do not depend on the transport coordinate $x$. Such an ideal
motor would consume one ATP molecule (with an energy amount of $0.5$ to $0.62$
eV or $20$ to $25$ $k_BT$ at room temperature, $k_B$ being the Boltzmann
constant) while transferring a cargo over
the distance $L$. The corresponding energy $E_{\rm in}(t)$ invested into 
the temporal increase
of the potential energy of the motor during repeating turnover cycles of the
`catalytic wheel'\cite{Wyman} can be calculated
as a sum of potential energy jumps $\Delta V(x(t_i))$ occurring at random instants of time
$t_i$ marking cyclic conformational 
transitions $V_1\to V_2\to V_1\to ...$
\cite{Sekimoto}.
Furthermore, apart from delivering a cargo over a distance $\delta x(t)$, the
motor can perform useful work, $W_{\rm use}(t)=f_0\delta x(t)$, against some
constant
force $f_0$ opposing its direction of motion. Ensemble averaging over many
trajectory realizations, we obtain the thermodynamic efficiency
$R_{\rm th}=\langle W_{\rm use}(t)\rangle/\langle E_{\rm in}(t)\rangle$. In the
long time limit this is a time-independent quantity in the normal transport
regime, where both $\langle E_{\rm in}(t)\rangle$ and $\langle W_{\rm use}(t)
\rangle$ are proportional to time. 
If $f_0=0$, the thermodynamic efficiency is
zero, and all the input energy will be dissipated as heat. In order to
characterize the energetic performance of a molecular motor in such a situation,
we introduce the delivery efficiency $D$ defined as the ratio of delivery
distance $d$ to the product of delivery time $t$ and the average number of
turnover cycles $\langle N_{\rm turn}\rangle$ (number of ATP molecules consumed).
$D$ thus has the meaning of a mean
delivery velocity per input energy amount (in dimensionless units). The goal is
to deliver cargo over a certain distance as quickly as possible using the
smallest amount of energy. Thus, ideally $d=L\; \langle N_{\rm turn}\rangle$
and $\langle N_{\rm turn}\rangle=\nu_{\rm turn}t$. Therefore, $D_{\rm ideal}=L^2
\nu_{\rm turn}/d$,
which linearly increases with the turnover frequency. However, we expect the
delivery efficiency to deviate from this idealization.

Both the motor and its cargo are subjected to friction and random thermal forces
from the environment. For normal viscous Stokes friction, the frictional 
drag force is $f_{
\rm fr}(t)=-\eta_0\dot{x}(t)$, where $\eta_0$ is the friction coefficient. It is
proportional to the medium's viscosity $\zeta$ and the particle size. For a
sphere of radius $a$, $\eta_0=6\pi a\zeta$. If we assume that the linker between
the motor and its cargo is rigid, we can model their coupled motion as that of a
point
particle moving under an effective frictional force. In this way, one accounts
for the cargo size simply by adjusting the effective friction. The dynamics for
such a simplified motor is then defined by the Langevin equation
\begin{equation}
\label{Langevin}
\eta_0\dot{x}(t)=-\frac{\partial V(x,t)}{\partial x}-f_0+\xi_0(t),
\end{equation}
where $\xi_0 (t)$ is a Gaussian random thermal force with zero mean,
which is completely characterized by its autocorrelation function. The
fluctuation-dissipation relation $\langle\xi_0(t)\xi_0(t')\rangle=2k_BT\eta_0
\delta(t-t')$ ensures that the description is compliant with the laws of
thermodynamics, so that no directional motion can emerge when $V(x,t)$
is time-independent and fixed to either $V_1(x)$ or $V_2(x)$. The prototype
motor model (\ref{Langevin}) has been investigated in great detail within
a Markovian setting, using various model potentials and a different number
of motor substates.

Our focus here is different, for the following experimental facts. As mentioned,
even in the absence of cargo the effective friction coefficient for the motor is
enhanced by a factor of 100 to 1000 in the cytosol compared with the one in
water\cite{Julicher2}. This phenomenon is due to the superdense state of the
cytosol, crowded with various biopolymers. Concurrently,
the diffusing nanoparticles themselves experience a medium with an
effectively enhanced viscosity, that depends on the particle size\cite{Luby}.
Moreover, numerous experiments reveal\cite{Guigas,Robert,Qian,Weigh} that the
complex shear modulus $G^*(\omega)$ of the cytosol displays a power law scaling
$G^*(\omega)\propto(i\omega)^\alpha$, $\alpha$ ranging between 0.2 and 0.9 for
frequencies in the  range from inverse milliseconds to several hundred inverse
seconds\cite{Yamada,Guigas,Robert,Mason, Weigh}. This reflects the viscoelastic
nature of the cytosol, which needs to be taken into account for molecular motors
\cite{Caspi,Robert,Bruno,Goychuk10,Goychuk12,GKh12a,KhG12,GKh13a,KhG13a}. To
explicitly consider viscoelastic effects for cargo particles
is even more pressing given the experimental results revealing
viscoelasticity-induced subdiffusion of submicron
particles\cite{Jeon,Caspi,Guigas,Tabei,Szymanski},
whose size is comparable to typical,
larger cargo such as vesicles.
This implies that: (i) the dynamics must
be described by a frequency-dependent
friction corresponding to a viscoelastic memory for the friction term with
a power-law kernel\cite{Goychuk12} $\zeta(t)\propto t^{-\alpha}$, such that
(ii) the diffusion of free particles becomes anomalously slow $\langle\delta
x^2(t)\rangle\propto t^{\alpha}$ with $0<\alpha<1$ on the corresponding time
scales from milliseconds to minutes. This is a mesoscopic, transient effect.
However, it becomes very important for transport processes in living cells as
they occur on exactly the physiologically relevant
time and length scales. Consequently, the Langevin
equation (\ref{Langevin}) for the motor's velocity must be extended to
the generalized Kubo-Langevin form
\begin{eqnarray}
\label{GLE}
\eta_0 \dot x(t)=&&-\frac{\partial V(x,t)}{\partial x}-f_0+\xi_0(t)\nonumber \\
&& -\int_{-\infty}^t\eta_m(t-t')
\dot x(t')dt'+\xi_m(t),
\end{eqnarray}
where $\eta_m(t)\propto\zeta(t)$, and 
$\xi_m(t)$ represents colored thermal Gaussian noise with autocorrelation
function $\langle \xi_m(t)\xi_m(t')\rangle=k_BT\eta_m(|t-t'|)$, as demanded by
thermodynamics (absence of directed  transport in a static periodic potential
$U_i(x)$) and the Kubo second fluctuation-dissipation theorem\cite{Kubo}. 
On physical grounds, a memory cutoff always
exists for $\eta_m(t)$ such that on a sufficiently long intermediate time
scale the kernel has the scaling property $\eta_m(t)\propto t^{-\alpha}$. This
ensures that the effective friction coefficient $\eta_0+\eta_{\rm eff}$ with
 $\eta_{\rm eff}=\int_0^{
\infty}\eta_m(t)dt$  is finite, but strongly enhanced over $\eta_0$,
in compliance with previous studies of molecular
motors\cite{Julicher2}. 
There exists also a short time cutoff,
corresponding to the largest vibrational frequency of the medium contributing
to the friction. Given the upper and lower cutoff, one can approximate the
memory kernel $\eta_m(t)$ by a finite  sum of
exponentials\cite{Goychuk09,Goychuk12}
and the generalized Langevin equation (\ref{GLE}) can be derived from
a corresponding multi-dimensional Markovian Maxwell-Langevin model of
viscoelastic dynamics\cite{Goychuk12}, see \textbf{Methods} for details. 
On the basis of this theoretically and experimentally well-founded approach
the results herein were obtained from numerical analysis. \\

\noindent \textbf{Results} \\

\textbf{Perfectly normal transport}. 
The first major surprise is that even carrying large cargo particles
like magnetic endosomes with radius about 300 nm our model
motor can operate by an almost
perfect power stroke mechanism in the normal transport regime, as demonstrated
in Fig.~\ref{Fig2}. For this, the binding potential amplitude
should be sufficiently large, $U_0=0.25$ eV, and the turnover rate sufficiently 
small, $\nu_{\rm turn}\sim 100$ Hz, both reasonable values
for this motor-cargo system. This almost perfect dynamics
occurs in spite of the fact that the free cargo alone
subdiffuses, $\langle\delta x^2(t)\rangle\sim2D_{\alpha}t^{\alpha}/\Gamma(1+
\alpha)$, on a transient time scale between $\tau_{\rm max}/\tilde\eta_{\rm
eff}^{1/(1-\alpha)}$ (see Methods)
and $\tau_{\rm max}$ with a subdiffusion coefficient as
small as $D_\alpha\approx3.68\times10^{-16}{\rm m^2/s^{0.5}}$ for $\tilde\eta_{
\rm eff}=10^4$ and $\tau_{\rm max}=22.4$ sec. This normal transport regime is
possible as viscoelastic subdiffusion is ergodic\cite{pt,Goychuk09,Goychuk12}.
Moreover, for a sufficiently large potential barrier $\Delta V$ separating
spatial periods of the potential the particle has enough time $\tau_{\rm rel}$
to relax and settle down sufficiently close to the potential minimum. Thus, it
can be advanced by a half-period at each switch of the potential, despite the
power-law character of the relaxation dynamics, if only the time scale
separation condition $\tau_{\rm rel}\ll \tau_{\rm turn}\ll\tau_{\rm esc}$ is
satisfied. This condition is easy
to fulfill for realistically small turnover frequencies $\nu$ and realistic
$\Delta V\sim15$ to $30\; k_BT_{\rm room}$ since the mean escape time grows
exponentially fast with the barrier height, $\tau_{\rm esc}\propto\exp(\Delta
V/k_BT)$. The occurrence of such normal transport is consistent with most
observations on molecular motors. 
Thus we accomplished our first goal, to explain the active normal transport
of particles, that otherwise subdiffuse when they are not
attached to a motor. However, the model motor considered here is actually
assumed to be a little too strong. It has a stalling force of about 15 pN,
almost twice the typical value, see below. Let us therefore consider a
somewhat weaker binding  potential and faster enzyme turnover rates
to see the origin of anomalous transport for such  a large cargo.

\begin{figure}
\includegraphics[height=5.cm]{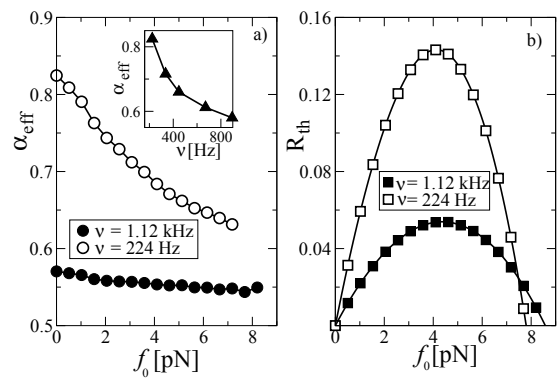}
\caption{Anomalous transport of large cargo particles at
lower potential amplitude, larger turnover rates, and
in the presence of loading force $f_0$.
(a) Effective anomalous transport exponent $\alpha_{\rm eff}$ and
(b) thermodynamic efficiency $R_{\rm th}$ while working against a constant
force $f_0$  near the end point of the simulations ($0.224$ sec or $10^6$
in dimensionless units). The thermodynamic efficiency decays over time as $R_{\rm
th}(t)\propto 1/t^{1-\alpha_{\rm eff}}$. The analysis considers the same
particles as in Fig.~\ref{Fig2}, but here the potential height is reduced
by factor of $2/3$. Ensemble averaging is performed over $10^3$ particles and
random realizations of potential flashes. The inset in (a) shows the dependence
of $\alpha_{\rm eff}$ on the mean enzyme turnover frequency for $f_0=0$.}
\label{Fig3}
\end{figure}

\textbf{Anomalous transport}. Indeed, when we decrease the potential amplitude
$U_0$ by $2/3$, increasing 
turnover frequency and in the presence of an opposing
(external drag) force $f_0$, which further reduces the effective barrier height,
an anomalous
transport regime $\langle\delta x(t)\rangle\propto t^{\alpha_{\rm eff}}$ with
$0<\alpha_{\rm eff}<1$ is enforced, see Fig.~\ref{Fig3},a. The thermodynamic
efficiency in this strongly anomalous transport regime is rather small, 
see Fig.~\ref{Fig3},b. Moreover, it becomes an algebraically decaying function
of time\cite{GKh13a,KhG13a}, $R_{\rm th}(t)\propto 1/t^{1-\alpha_{\rm eff}}$.
This happens because $W_{\rm use}(t)\propto\langle\delta x(t)\rangle\propto
t^{\alpha_{\rm eff}}$ scales sublinearly with time, while the input energy
is consumed at a constant rate. This means that asymptotically most of the
input energy is used to overcome the dissipative influence of the environment
characterized by the massively enhanced effective viscosity. Concurrently, the
useful
work performed against the force $f_0$ always remains finite and the stalling
force $f_0^s$ is also about the same as for normal diffusion ratchets. This finding
agrees with typical experimental values of 7 to 8 pN for
the stalling force of kinesin
motors\cite{Julicher2}. The slower the motor turnover is, the larger the
effective transport exponent $\alpha_{\rm eff}$ becomes, along with a higher
motor efficiency. The efficiency displays almost a parabolic dependence on
the force $f_0$, $R_{\rm th}\propto f_0(1-f_0/f_0^s)$ in this regime.
A similar dependence was derived for fluctuating
tilt viscoelastic ratchets\cite{GKh13a,KhG13a}. The
stalling force is not only roughly
proportional to the barrier height, but also depends on the frequency of the
potential switches.  

\begin{figure}
\includegraphics[height=5.7cm]{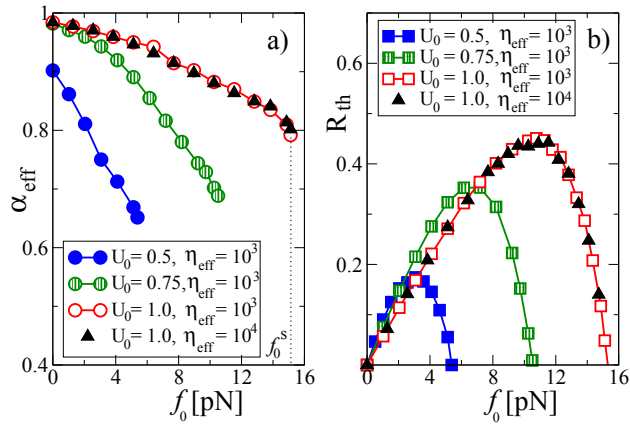}
\caption{
Dependence of (a) the effective transport exponent $\alpha_{\rm eff}$ and (b)
the thermodynamic efficiency $R_{\rm th}$ on the load $f_0$, for three different
potential amplitudes and for turnover frequency $\nu_{\rm turn}=1.12$ kHz. 
Ensemble averaging is done over $10^3$ particles
and random realizations of potential flashes, $\tilde \eta_{\rm eff}=10^3$,
$\tau_{\rm max}=22.4$ sec, or $\tilde \eta_{\rm eff}=10^4$ and  $\tau_{\rm
max}=2240$ sec (with the same $\eta_{\alpha}$, see \textbf{Methods}). Matching
of the results for two sets with the same amplitude $U_0=1.0$ indicates that
$\eta_{\alpha}$ is the characteristic quantity, rather than
$\eta_{\rm eff}$ and  $\tau_{\rm max}$ separately. Efficiency is calculated
at the end point of simulations $10^6$ corresponding to $t=0.224$ sec.}
\label{Fig4}
\end{figure}

\textbf{Smaller cargo size.} Depending on the cargo
size and the binding potential amplitude the transport can become more normal
and thermodynamically highly efficient even for a large
turnover frequency, as Fig. \ref{Fig4} illustrates for $\nu_{\rm turn}=1.12$ kHz.
This is about the maximal turnover frequency which
can be expected for molecular motors.
For this, the potential amplitude
should be sufficiently large and the cargo smaller in size. Here,
we reduced  $\tilde \eta_{\rm eff}$ to $10^3$ with the same $\tau_{\rm
max}=22.4$ sec. Hence, assuming that the effective viscosity of the medium remains
unchanged, $\eta_{\alpha}\sim\eta_{\rm eff}\tau_{\rm max}^{\alpha-1}$ becomes
reduced by a factor of ten, which corresponds to a cargo with one tenth of the
size, that is, of some 30 nm radius. However, if we were taking into account its
dependence on the particle size\cite{Luby}, this value of $\tilde\eta_{\rm eff}$
should in fact be attributed to particles with approximate radius of 100 nm.
The subdiffusion coefficient is enhanced
accordingly, $D_\alpha \approx 3.68\cdot 10^{-15}\;{\rm m^2/s^{0.5}}$.
Furthermore, to show that the fractional friction strength $\eta_{\alpha}$
and the subdiffusion coefficient $D_\alpha$ are characteristic for the transport
properties rather than $\eta_{\rm eff}$ and $\tau_{\rm max}$ separately, we
also considered the case with $\tilde \eta_{\rm eff}=10^4$ and $\tau_{\rm
max}=2240$ sec yielding the same $D_\alpha$, see data with $U_0=1.0$ in
Fig.~\ref{Fig4}.
For the largest potential
amplitude in Fig.~\ref{Fig4} the thermodynamic efficiency is appreciably high,
up to $45$\% for the studied case. This is very surprising: The transport
efficiency in the anomalous regime can be temporally almost as high as the
maximal efficiency of kinesin motors in the normal regime (about 50\%).
For this potential amplitude, however, our motor is stronger than a typical
kinesin motor. It has a stalling force of about 15 pN, see Fig.~\ref{Fig4}.
The effective exponent $\alpha_{\rm eff}$ is about $0.87$ at this maximum.
However, the transport is anomalous and the efficiency decays algebraically as
$R_{\rm th}(t)\propto t^{-0.13}$. For increasing loading force $f_0$ the
anomalous diffusion exponent $\alpha_{\rm eff}$ becomes smaller and the
thermodynamic efficiency drops faster as function of time.
This means that the optimal value of force $f_0$ corresponding to the maximum
of $R_{\rm th}$ slowly shifts towards smaller values, as if the motor became
gradually ever more tired, and more quickly exhausted for a higher load.
Upon reduction of the barrier height by a factor of $3/4$ the transport is still
close to normal for $f_0=0$. Moreover, the thermodynamic efficiency can
still be temporally rather high at optimal load. However,
$\alpha_{\rm eff}$ drops now faster with $f_0$. A stronger reduction of
the potential amplitude, by one half, leads immediately to the emergence of a 
low efficiency, strongly anomalous transport regime, even for $f_0=0$
(Fig.~\ref{Fig4}). However, a strong reduction of the turnover frequency 
down to $\nu_{\rm turn}=100$ Hz will recover the normal transport regime for a sufficiently
small $f_0$ in all cases considered. The transport of even smaller particles
is clearly normal for realistic parameters.

\begin{figure}
\resizebox{0.8\columnwidth}{!}{\includegraphics{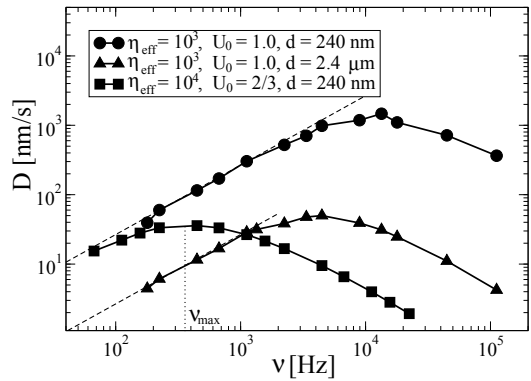}}
\caption{Delivery efficiency $D$ as function of the turnover frequency $\nu_{
\rm turn}$. In the ideal power-stroke regime, $D_{\rm ideal}=L^2 \nu_{\rm turn}/d$. For
small $\nu_{\rm turn}$ our results agree with this simple dependence (broken
lines).}
\label{Fig5}       
\end{figure}

\textbf{Delivery performance.} For vanishing loading force $f_0=0$ thermodynamic
efficiency is zero, even in the normal transport regime. We proposed above that
the performance of molecular motors such as kinesin should be characterized by
the energetic efficiency $D$ of the cargo delivery over a certain
distance $d$.
We calculated the above-defined delivery
efficiency $D$ in Fig.~\ref{Fig5} as function of the turnover frequency 
$\nu_{\rm turn}$ for several different delivery distances $d$, barrier heights
$U_0$, and 
fractional friction
coefficients $\eta_{\alpha}$. Remarkably, for a small cargo (smaller $\eta_{\rm
eff}$ in Fig.~\ref{Fig5}) the calculated delivery efficiency follows the ideal
power-stroke dependence, $D=L^2\nu_{\rm turn}/d$ in 
the entire range of realistic turnover frequencies.  Even for a relatively large
cargo, Fig.~\ref{Fig3}, but for much smaller turnover frequencies the transport
is close to this ideal normal regime. It is expected to be normal
already for $\nu_{\rm turn}=10$ Hz. Such a small frequency is, however, not
accessible to numerical analysis. In the anomalous regime, with increasing
turnover frequency the delivery efficiency reaches a maximum at $\nu_{\rm max}$
and then decreases. The anomalous transport becomes less efficient for 
$\nu>\nu_{\rm max}$ and $\nu_{\rm max}$ shifts to smaller values with increasing
delivery distance $d$. We speculate that in the cell economy this effect can
become relevant for the optimization of the metabolic budget.\\

\noindent \textbf{Discussion}\\

We showed that transport by molecular motors becomes anomalous for large 
cargo particles with large fractional friction coefficient  $\eta_\alpha$ 
 when the enzyme turnover is fast, and the binding 
potential amplitudes $\Delta V$ is not sufficiently large. Larger potential
amplitude $\Delta V$ for the fixed spatial period leads to faster relaxation
of the motor particle to a new potential minimum after each potential switch.
The enhancement
of $\eta_\alpha$ leads to slower relaxation which
asymptotically follows a power law decay $t^{-\alpha}$. For this reason,
to guarantee the occurrence of normal transport the condition 
$\nu_{\rm turn}^{-1}\gg \tau_{\rm rel}$ must be fulfilled, the separation
of time scales should account to several orders of magnitude. 
Since the potential curvature can be estimated as $\kappa=(2\pi/L)^2\Delta V$,
and the anomalous relaxation rate as\cite{Goychuk09} $\tau_{\rm rel}\sim(\eta_{
\alpha}/\kappa)^{1/\alpha}$, one can see that the ratio $\eta_{\alpha}/\Delta V$
is important to determine the scale of $\tau_{\rm rel}$. Even if the transport
of a large cargo is anomalous for $\Delta V\approx 0.47$ eV in Fig.~\ref{Fig3}
the reduction of $\eta_{\alpha}$ rapidly enforces normal transport, similarly
to the reduction of $\nu_{\rm turn}$. Therefore, for realistic turnover
frequencies the molecular transport by kinesins is expected to be normal for
vesicles of a typical radius of 30 nm, and possibly up to 100 nm. 
Anomalous transport emerges for
large cargos with radius $\sim 300$ nm or larger. This explains why
the same molecular motors can mediate both normal and anomalous transport
in living cells 
depending on the cargo size. 
The occurrence of an anomalous transport exponent is thus reconciled with the
the normal transport behavior for small cargo at lower turnover frequencies.

Our research provokes a number of followup questions. Thus, what happens
if we relaxed the assumption of a rigid motor-cargo linker molecule? In that
case, the large subdiffusing cargo is elastically coupled to a molecular motor,
that possibly still operates normally in the absence of cargo. We are currently
investigating  this generalization for realistic spring constants of the linker.
However, qualitatively the results remain very similar. Another question is
prompted by the experimental results in Ref.~\cite{Robert}, suggesting that
the motors can collectively transport several magnetosomes jointed into a chain.
A generalization of our non-Markovian model to such collectively operating motors
would be important for our understanding of large-cargo transport in living
cells.

Concluding, we proposed a simple basic model which reconciles experimental
observations of both normal and anomalous transport by identical, highly
processive molecular motors in biological cells. Our model presents an immediate
generalization of a well-known two-state model of normally diffusing molecular
motors accounting for the viscoelastic properties of the intracellular fluid.
It not only explains how molecular motors may still operate by a
power-stroke like mechanism while carrying a large cargo which subdiffuses
when left alone, but also why and how an anomalous transport regime emerges
for even larger cargo. It thus presents a good starting point for future
research and further generalizations. Understanding how molecular motors
perform in the viscoelastic cytosol of living cells despite the subdiffusion
of the free cargo is compelling. Our findings open new vistas to the old problem
of intracellular trafficking, reconciling seemingly conflicting results for the
motor-cargo dynamics under different conditions. Finally, our results will be
of crucial importance for the design of new technologies of motor-driven particle
and drug delivery in the crowded cytosol of cells. We are confident that our
findings will prompt a series of new experiments on the dynamics of molecular
motors under realistic conditions in live cells.
\\
\vspace{1cm}

\noindent \textbf{Methods} \\

\small

The numerical approach to integrate the dynamic equation (\ref{GLE}) with a
power-law rests on the approximation of the memory kernel by a sum of
exponentials\cite{Goychuk09,Goychuk12}, 
\begin{eqnarray}
\label{memory}
\eta_m(t)=\sum_{i=1}^{N}k_i\exp (-\nu_it).
\end{eqnarray}
The rate constants $\nu_i$ and elastic constants $k_i$ are chosen to obey a
fractal scaling\cite{Palmer}, $\nu_i=\nu_0/b^{i-1}$, $k_i\propto \nu_i^\alpha$,
with a dilation parameter $b>1$, and a shortest memory time $\nu_0^{-1}$ in
the hierarchy. Due to the scaling property $\eta_m(ht,\nu_0)=h^{-\alpha}\eta_m
(t,h\nu_0)$ this choice indeed provides a power law regime, $\eta_m(t)\propto
t^{-\alpha}$ on time scales $\nu_0^{-1}\ll t\ll\tau_{\rm max}=b^{N-1}/\nu_0$,
with small superimposed logarithmic oscillations. Physically, this corresponds
to representing a viscoelastic environment by auxiliary Brownian quasi-particles
with coordinates $x_i$. They are coupled to the central Brownian particle by
elastic constants $k_i$ and are subject to the thermal noises and frictional
forces with viscous frictional constants $\eta_i=k_i/\nu_i$\cite{Goychuk12},
\begin{eqnarray}
\label{embedding}
\eta_0 \dot x & =& f(x,t)-
\sum_{i=1}^{N}k_i(x-x_i)+\xi_0(t) \;,\nonumber \\
\eta_i\dot x_i& = &k_i (x-x_i)+\sqrt{2\eta_ik_BT}\xi_i(t) \;,
\end{eqnarray}
where $f(x,t)=-\partial U(x,t)/\partial x$, and $\xi_i(t)$ are uncorrelated
white Gaussian noises of unit intensity, $\langle\xi_i(t')\xi_j(t)\rangle=
\delta_{ij}\delta(t-t')$, which are also uncorrelated to $\xi_0(t)$. To have a
complete equivalence with equation (\ref{GLE}), the initial positions
$x_i(0)$ are sampled from
a Gaussian distribution centered around $x(0)$, $\langle x_i(0)\rangle=x(0)$
with variance $\langle[x_i(0)-x(0)]^2\rangle=k_BT/k_i$. We set
\begin{eqnarray}
k_i=\nu_0\eta_{\rm eff }\frac{b^{1-\alpha}-1}{b^{(i-1)\alpha}[b^{N
(1-\alpha)}-1]}
\end{eqnarray}
and use $b=10$ which leads to a maximal relative error with respect to the
exact power-law of less than 4\% \cite{GKh13a}, for $\alpha=0.5$, on relevant
intermediate time scale. The effective relative friction coefficient
$\tilde \eta_{\rm eff}= \eta_{\rm eff}/\eta_0\gg 1$ is used as a parameter in
our simulations. The parameter $\nu_0$ controls the matching accuracy
of our model with the model of fractional Langevin dynamics (the
memory kernel $\eta_m(t)=\eta_\alpha t^{-\alpha}/\Gamma(1-\alpha)$) at
short to intermediate times, where initially for times $t\ll \tau_{\rm
in}=(\eta_0/\eta_\alpha)^{1/(1-\alpha)}$ the free diffusion is normal,
$\langle \delta x^2(t)\rangle \sim 2 D_0 t$ with $D_0=k_BT/\eta_0$ 
in accordance with the Einstein-Stokes relation. Then, anomalous
diffusion emerges, $\langle \delta x^2(t)\rangle \sim 2 D_\alpha t^\alpha/
\Gamma(1+\alpha)$ with $D_\alpha=k_BT/\eta_\alpha$. The number $N$ 
of auxiliary quasi-particles controls the maximal
range of subdiffusive dynamics, which becomes again normal,
$\langle \delta x^2(t)\rangle \sim 2 D_{\rm eff} t$, for  $t\gg
\tau_{\rm max}= b^{N-1}/\nu_0$ with $D_{\rm eff}=k_BT/(\eta_{\rm
eff}+\eta_0)$. Note that the fractional viscosity $\eta_{\alpha}=\eta_{\rm eff}\tau_{\rm
max}^{\alpha-1}/r$, where $r=\frac{C_\alpha(b)}{\Gamma(1-\alpha)}
\frac{b^{1-\alpha}}{b^{1-\alpha}-1}[1-b^{-N(1-\alpha)}]$, is a
numerical coefficient, $r\approx 1.07$ for $N\geq 6$,
$\alpha=0.5$, and $b=10$, with $C_\alpha(b)\approx 1.3$\cite{Goychuk09}. 
An interesting observation is that in terms of $\tau_{\rm max}$ and 
$\tilde\eta_{\rm eff}$, $\tau_{\rm in}\sim \tau_{\rm max}/\tilde\eta_{\rm
eff}^{1/(1-\alpha)}$. Therefore, the effective viscosity $\tilde\eta_{\rm
eff}$ defines the time range of subdiffusion, from $\tau_{\rm
max}/\tilde\eta_{\rm eff}^{1/(1-\alpha)}$ to $\tau_{\rm max}$,
independently of $b$, and $N$! For example, for $\tilde\eta_{\rm eff}=10^3$
and $\alpha=0.5$ one expects that subdiffusion will extend over 6 decades in
time. In the simulations, we scale length in units of $L$ and time in units
of $\tau_0=L^2\eta/(4\pi^2 U_0)$, where $U_0$ was taken equal to $U_0=0.25$
eV (or about 10 $k_BT_{\rm room}$ with fixed temperature). The
time step was $\delta t=0.01$ and $t_{\rm max}$ was varied from $10^6$ to
$10^7$. The ratio $U_1/U_0$ was fixed to $0.65$. 
Four different values of $U_0$ were used: $U_0=1$, $3/4$, $2/3$, and $1/2$.
The largest one corresponds to the largest potential barrier $\Delta V$ separating
two potential periods of about $28\;k_BT_{\rm room}\approx 0.7$
eV, $2/3$ to about  $18.7\; k_BT_{\rm room}\approx 0.47 $ eV, and the
smallest one to $14\;k_BT_{\rm room}\approx 0.35$ eV. The larger $U_0$
the more efficiently the motor works, as the probability of thermally activated
backsteps is exponentially suppressed with $\Delta V/k_BT$. This is necessary to
provide an ideal power-stroke operation at small turnover frequencies.
However, the energy released from the hydrolysis of one ATP molecule will
not be sufficient to perform one cycle of operation in the case $\Delta V_0=0.7$
eV because of the energy derived from the hydrolysis of one ATP molecule 
does not exceed $0.62$ eV. To drive a cycle of two potential flashes one
needs an input energy of at least 0.7 eV (0.35 eV per one conformational
change, about one half of potential barrier separating potential periods, 
see upper inset in Fig. \ref{Fig2}). 
The power exponent of anomalous diffusion was fixed to $\alpha=0.5$ to
interpolate between $\alpha=0.4$ for an intact cytoskeleton\cite{Robert,Bruno}
and $\alpha=0.56$\cite{Robert} when the actin filaments are disrupted. For the
radius $a=300$ nm of a single magnetic endosome\cite{Robert} one estimates 
$\eta_0\approx 5.65 \cdot 10^{-9}\;{\rm N \cdot s/m}$ in water of viscosity
$\zeta=10^{-3}\;{\rm Pa \cdot s}$. This yields  $\tau_0\approx 2.24
\cdot 10^{-7}\;{\rm s} $. We use $\nu_1=\nu_2$ in the simulations. Hence,
$\nu=\nu_1/2$, and the frequency $\nu_1=5\cdot 10^{-4}$ corresponds to $\nu
\approx 1.12 $ kHz. This is about the maximal possible turnover frequency
for molecular motors\cite{Julicher2}.  For an effective particle ten times
smaller, $\tau_0\approx 2.24 \cdot 10^{-8}\;{\rm s}$ and $\nu_1=5\cdot
10^{-5}$ would correspond to the same physical turnover frequency. For a
magnetic endosome we estimate $\tilde\eta_{\rm eff}=10^4$, $\nu_0=0.1$
and take $N=8$, which yields $\tau_{\rm max}\approx 22.4 $ sec and
$D_\alpha \approx 3.68\cdot 10^{-16}\;{\rm m^2/s^{0.5}}$, of the same order
of magnitude as in experiments for a cargo consisting of several such
endosomes\cite{Robert}. Alternatively, with a smaller $\tilde\eta_{\rm eff}=10
^3$ for $\nu_0=1$ and $N=9$, $D_\alpha$ is by a factor of ten larger,
$D_\alpha \approx 3.68\cdot 10^{-15}\;{\rm m^2/s^{0.5}}$, at the same
$\tau_{\rm max}$. One obtains the same $D_\alpha$ by simultaneously increasing
$\tilde\eta_{\rm eff}$ to $\tilde\eta_{\rm eff}=10^4$ and $\tau_{\rm max}$
to  $\tau_{\rm max}=2240$ sec ($N=11$).

\textbf{Acknowledgments}\\
Support of this research by the German Research Foundation, Grants
GO 2052/1-1 and GO 2052/1-2, as well as funding from the Academy of Finland
(FiDiPro scheme) are gratefully acknowledged. We also thank our colleagues
Andrey Cherstvy and Aleksei Chechkin for useful discussions.\\

\textbf{Author contributions} \\
I.G., V.O.K, and R.M. designed the research; I.G. and V.O.K. carried out the
research; I.G., V.O.K and R.M. wrote the paper.\\

\textbf{Additional information} \\
Supplementary Information accompanies this paper on
www.nature.com/naturephysics.\\

\textbf{Competing financial interests} \\
The authors declare no competing financial interests.

\end{document}